# Assessing Attendance by Peer Information

Pan Deng[1], Jianjun Zhou[1*], Jing Lyu[2], Zitong Zhao[2]
[1]Shenzhen Research Institute of Big Data, Shenzhen, China
[2]The Chinese University of Hong Kong, Shenzhen, China
{pandeng, jinglyu, zitongzhao}@link.cuhk.edu.cn
zhoujianjun@cuhk.edu.cn

## ABSTRACT

Attendance rate is an important indicator of students' study motivation, behavior and Psychological status; However, the heterogeneous nature of student attendance rates due to the course registration difference or the online/offline difference in a blended learning environment makes it challenging to compare attendance rates. In this paper, we propose a novel method called Relative Attendance Index (RAI) to measure attendance rates, which reflects students' efforts on attending courses. While traditional attendance focuses on the record of a single person or course, relative attendance emphasizes peer attendance information of relevant individuals or courses, making the comparisons of attendance more justified. Experimental results on real-life data show that RAI can indeed better reflect student engagement.

## 1. INTRODUCTION

While studying offline is the norm for most schools, during epidemic periods all or a portion of students are forced to study online due to university closure. In such a blended learning environment, tracking the study status and the wellbeing of students is an important issue for the university. Students' attendance in classes is a measure that reflects students' enthusiasm for the course and their status in the university [29]. Many studies suggest a correlation between attendance and attainment at university [5, 26, 14]. Several studies detect attendance rates using mobile devices and include attendance as a feature to predict academic performance [27, 19, 28]. Attendance is also correlated with behavior and Psychological problems such as video game addiction [24] and depression [26]. Detecting unusual attendance rate changes can help to identify abnormal behaviors and Psychological problems in an early stage and provide in time intervention to students in need.

The successful applications of attendance data call for fair comparisons among peer attendance, especially in universities.

Traditionally, the attendance rate of a course or a student is isolated and might not be compared fairly, because in many universities students are allowed to select some courses on their own so that the course registration records of two students can be different. In addition, each course can have its own attendance policy, making it harder to compare attendance rates. Courses that have mandatory attendance requirements usually have higher attendance rates than those do not, so that it is not fair to compare course attendance rates without considering attendance requirements. Similarly, students who registered for courses with mandatory attendance requirements usually have higher attendance rates than students who registered for courses with voluntary attendance requirements, so that the attendance rate does not always reflect the attainment of a student. Attendance of online and offline courses may not be compared directly as well, because the efforts to attend those courses can be significantly different. Attending online courses could be as simple as a mouse click away, while attending offline courses usually requires travelling from place to place physically.

Traditionally it is not easy to fairly compare attendance in a university, due to not just the diversity of course registration and attendance requirements but also the difficulty of collecting campus-wide attendance data. Without attendance information of peer students or courses, the attendance data of a student or a course is isolated and difficult to adjust. However, in the era of Big Data, many new technologies [12, 18, 31] have been proposed to collect attendance data for many courses simultaneously, making it possible to analyze the attendance structure of the student population, and develop new attendance calculation methods.

Careful comparisons of attendance can also provide insight into students' academic interest. If a student attends a course that has a generally low attendance rate, it indicates that the student is more willing to attend the course than their classmates are; On the other hand, if a student attends a course that has a generally high attendance rate, it indicates that the student is just doing what others are doing.

In this paper, we propose a novel method called Relative Attendance Index (RAI) to measure attendance, which reflects the efforts on attending courses and makes comparisons of attendance more justified. To our knowledge, this is the first study on fair comparisons of attendance. While traditional attendance focuses on the record of a single person or course, we define a notion for attendance contribution to course attendance and add the attendance information of relevant individuals or courses to make the comparisons of attendance fairer. We perform a campus-wise study on attendance and analyze its effects on course grades and GPA. Our experiment results show that RAI has a higher correlation with academic performance than the traditional attendance rate.

The rest of this paper is organized as follows. Section 2 describes the related studies. Section 3 introduces the RAI definition. Section 4 presents the experiment results on real-life data from a university. Section 5 discusses an application of RAI on clustering student populations. Section 6 describes the limitations and future work. Section 7 lists the acknowledgments.

## 2. RELATED STUDIES

With the advancement of technology, many new methods have been proposed to collect campus-wide attendance data. Several studies [13, 11, 18] measured attendance via QR code systems in which QR codes are generated and then scanned by students to authenticate themselves. Wang et al. [26] deployed an APP to students' cell phones to detect attendance by GPS signals and WiFi tracing. A method independently developed in [19] and [28] used WiFi log to calculate attendance. Studies in [2, 25] proposed Bluetooth/Beacon based attendance prediction systems. Shoewu and Idowu [22] used fingerprints and Kar et al. [12] used face recognition to detect individual attendance rates.

Some studies measured offline and online attendance at the same time. Brennan et al. [4] detected physical attendance by thermal sensors and online behaviors by clickstream data. The change of online and physical attendance through time was observed. However, due to a technology limit, the method did not link physical attendance to individuals and did not study the issue of attendance comparison. Nordmann et al. [17] mixed the data of physical attendance and online recording clickstream together to form the total attendance rate instead of studying them separately. However, attendance of live lectures is still a stronger predictor than recording use on students' academic performance.

Many studies confirmed the correlation between attendance and academic performance [1, 3, 7, 16]. See [15] for a survey. [15] also reviewed factors that affect attendance. To work around the issue of fair comparisons of attendance, many studies focused on samples from the same course or samples with similar registration records (e.g., first year students) [1, 3, 10]. [3] also controlled factors such as age, gender, nationality etc. in their regression analysis. Studies in [7] and [16] divided the students into bands according to grades and used the average attendance of each band for correlation studies.

Student subtyping and clustering are widely used in analyzing learning process and predicting academic performance. Yang et al. [30] applied EM-IRL to students learning behavior data and observed significant differences between groups. Romero et al. [20] used clustering on online forum data to predict students' final performance. Resulting model turned out to be suitable and highly interpretable. Cerezo et al. [5] studied both learning process and clusters' relation with performance using LMS logs data. Resulting clusters are well-interpreted and showed satisfying correlation with final marks.

Many studies explored the reasoning for student class attendance. Friedman et al. [9] and Moore et al. [14] reported positive relationship between class attendance and students' motivation. Sloan et al. [23] further found that the level of interest has significant impact on attendance. These studies indicated that attendance, along with other features, can better show students' academic interest than traditional models.

None of the above studies has applied peer information to revise attendance measurements.

## 3. METHOD

The traditional attendance rate of a class or a student is defined in a straightforward way. Only the information of the class or the student is involved. We give the definition of Attendance Rate (AR) formally as in Definition 1.

**Definition 1 (Attendance Rate $r_c$ and $r_s$):** Given class $c$ and student $s$, let $n_c^{reg}$ and $n_c^{att}$ be the number of students registered $c$ and the number of students attended $c$ respectively; let $n_s^{reg}$ and $n_s^{att}$ be the number of classes registered by $s$ and the number of classes attended by $s$. Then the Attendance Rate (AR) of class $c$ ($r_c$) and of student $s$ ($r_s$) are defined as below respectively.

$$r_c = \frac{n_c^{att}}{n_c^{reg}}, \tag{1}$$

$$r_s = \frac{n_s^{att}}{n_s^{reg}} \tag{2}$$

Students can have different sets of registered classes, and classes can have very different attendance requirements. When comparing the attendance rates of two students, it is necessary to analyze the set of classes attended by these two students and the attendance rates of these classes. If a student attends a class attended by almost everyone, the student makes little contribution to the attendance rate of the class; on the other hand, if a student attends a class that has a low attendance rate, the student makes a significant contribution to the attendance rate. To capture the concept, we propose the notion of attendance contribution as in Definition 2.

**Definition 2 (Attendance Contribution $D_{sc}$):** Let $r_c$ be the attendance rate of class $c$, and $a_{sc}$ be a function indicting whether student $s$ attended class $c$ or not, then the Attendance Contribution of student $s$ on the attendance rate of class $c$ is defined as

$$D_{sc} = a_{sc} - r_c, \text{ with} \tag{3}$$

$$a_{sc} = \begin{cases} 1, & \text{if } s \text{ attended } c \\ 0, & \text{if } s \text{ did not attend } c \end{cases} \tag{4}$$

Since $r_c \in [0, 1]$, Attendance Contribution is a number between -1 and 1. If $s$ has registered $c$ and $s$ attended $c$, then the attendance rate of $c$ cannot be zero and $D_{sc}$ can approach 1 but never reach 1.

With Attendance Contribution, we can compare the attendance rates of two students by computing the average Attendance Contribution on registered classes. We defined the notion as Relative Attendance Index (RAI) in Definition 3.

**Definition 3** (Relative Attendance Index $RAI_s$): Given student $s$, Let $K_s$ be the set of classes registered by $s$, the Relative Attendance Index (RAI) of $s$ is defined as

$$RAI_s = \frac{\sum_{c \in K_s} D_{sc}}{|K_s|} \tag{5}$$

RAI considers both the student's individual attendance status of a semester and the attendance status of the student's classmates. The peer information is injected into the new measure through the course attendance rate in Attendance Contribution.

**LEMMA 1**: $-1 < RAI_s < 1$.
Proof: The $RAI_s$ definition only considers classes registered by $s$. When $a_{sc} = 0$, $r_c \in [0,1)$ ; When $a_{sc} = 1$, $s$ attended $c$, therefore $r_c \in (0,1]$ . Thus $a_{sc} - r_c \in (-1,1)$, Therefore $RAI_s = \frac{\sum_{c \in K_s}(a_{sc}-r_c)}{|K_s|} \in (\frac{|K_s|\times -1}{|K_s|}, \frac{|K_s|\times 1}{|K_s|}) = (-1, 1)$. □

RAI is a number between -1 and 1. When RAI approaches -1, $a_{sc}$ is mostly 0 and $r_c$ approaches 1 for most classes, indicating that the student has skipped many well attended classes. On the other hand, when RAI approaches 1, $a_{sc}$ is mostly 1 and $r_c$ approaches 0 for most classes, indicating that the student has attended many poorly attended classes. Therefore, RAI shows the difference of attitudes toward classes between students and their classmates.

## 4. Results
### 4.1 Data and Setup

The anonymous attendance and grade data used in this paper were collected in 2018 and 2019 from a university[1] in China. We applied the method proposed in [19] and [28] to calculate the attendance. Note that our Relative Attendance Index can be applied to attendance data collected using other methods such as QR code [18]. The student IDs were converted into hash codes, then the attendance and grade data were connected through the hash codes. The university did not have a mandatory attendance policy; however, while most instructors followed the university policy, some instructors had their own attendance requirements. Some instructors used in-class discussions and quizzes to encourage attendance. The data came from 4838 students from Cohort 14 to 19 in 44 majors, spanning over 3 semesters, with 1489 courses grouped into 37 categories by the university. For most courses, students received letter grades from A to F. Courses with other grades such as P/F were excluded. The traditional attendance rate (AR) for a student was calculated using Formula (2) in Definition 1, and the corresponding RAI was calculated using Formula (5) in Definition 3.

### 4.2 Correlation with Academic Performance

Many previous studies show that attendance is correlated with academic performance. Given that the purpose of student attendance comparison is usually to assess the attainment of the students, we calculated the correlation between attendance rates and GPA to assess the fairness of attendance comparisons. A more correctly calculated attendance assessment will have a higher correlation with the GPA. The Pearson correlation between RAI and GPA is 0.48, which is significantly higher than that of AR (0.37). The p-values of the two correlation values are $3.7 \times 10^{-225}$ and $2.6 \times 10^{-129}$ respectively. Since they are well below the 0.05 threshold, the correlation values are generally considered significant.

We also calculated the correlation between attendance and academic performance within each course category. The result is shown in Table 1 (sorted by the RAI correlation). Some course categories were filtered out because they had small enrollments and did not generate correlation values with low enough p-values (<0.05) to be statistically significant. For 19 out of the 26 course categories, RAI has a higher correlation than AR. Only for two categories, GED and FRN, AR has a higher correlation than RAI (The descriptions of the categories are listed in Table 1). AR and RAI are tie for the five categories of FMA, GNB, ERG, CHM and CSC. We remark that language related courses such as ENG (English) have low correlations because those courses usually have in-class discussions resulting in an unofficial mandatory attendance requirement. Categories that rely on prior knowledge in high school, such as Chemistry and Physics, also have low correlations. Since most students attending this university did not study Calculus in high school and Calculus accounts for a large faction in Mathematics courses, it is reasonable to see the MAT category having a much higher correlation.

**Table 1. Correlation in course categories**

| CAT. | Description | AR | RAI |
|---|---|---|---|
| FMA | Financial Mathematics | **0.65** | **0.65** |
| MSE | Material Science and Engineering | 0.46 | **0.52** |
| BIM | Bioinformatics | 0.35 | **0.51** |
| GED | General Education D | **0.48** | 0.46 |
| GEB | General Education B | 0.42 | **0.45** |
| STA | Statistics | 0.32 | **0.39** |
| MGT | Management | 0.26 | **0.39** |
| GFN | GE Foundation: In Dialogue with Nature | 0.28 | **0.37** |
| FIN | Finance | 0.34 | **0.36** |
| MAT | Mathematics | 0.34 | **0.36** |
| EIE | Electronic Information Engineering | 0.34 | **0.36** |
| GFH | GE Foundation: In Dialogue with Humanity | 0.34 | **0.36** |
| ECO | Economics | 0.27 | **0.36** |
| GNB | Genomics and Bioinformatics | **0.35** | **0.35** |
| GEA | General Education A | 0.32 | **0.35** |
| ACT | Accounting | 0.32 | **0.34** |
| PHY | Physics | 0.22 | **0.29** |
| HSS | Humanities and Social Science | 0.27 | **0.28** |
| ERG | General Engineering courses | **0.27** | **0.27** |
| GEC | General Education C | 0.24 | **0.27** |
| CHM | Chemistry | **0.25** | **0.25** |
| FRN | French | **0.28** | 0.23 |
| CSC | Computer Science | **0.23** | **0.23** |
| MKT | Marketing | 0.12 | **0.22** |
| CHI | Chinese | 0.13 | **0.19** |
| ENG | English | 0.06 | **0.12** |

### 4.3 RAI Distribution

To illustrate the different distributions on RAI for high and low course grade students, we collected two sets of samples, with one set having a course grade no less than B+ and the other set no greater than C. Each sample is a triplet with a hashed student ID, a course ID, and the corresponding grade received by the student in the course. We then calculated the RAI of the student in the corresponding course. Figure 1 (a) shows the distribution of the first set. It shows that more than 50% of samples have RAI > 0 (better than normal). Figure 1 (b) shows the distribution of the low course grade set. It shows that the majority of samples have RAI < 0 (worse than normal), with some down to -0.8. For easier comparisons of both sets, the values in both subfigures have been normalized as the proportion values. We can see that the first set has a more concentrated distribution than the second set. This indicates that students receiving grade C or lower have a much higher probability of having extreme attendance behaviors (skipping many courses).

---

[1] The use of the data by our project has been approved by the university management and the committee in charge of personal information in this university.

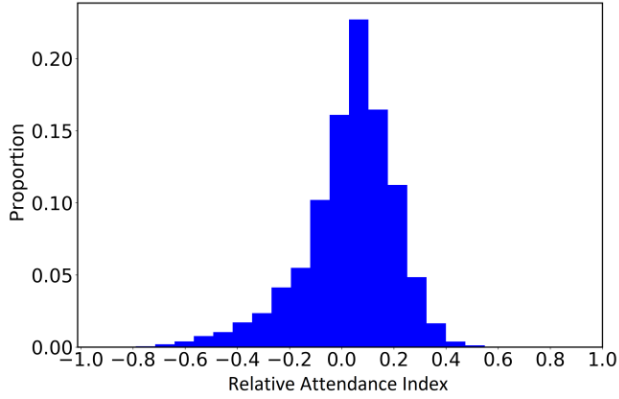

(a) Samples with grades ≥ B+

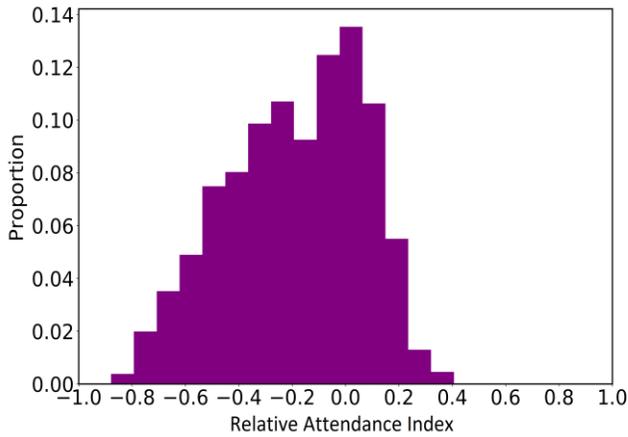

(b) Samples with grades ≤ C

**Figure 1. RAI of high and low course grade samples.**

## 5. DISCUSSION

In this section, we showcase an application of RAI on clustering the student population.

We formatted the attendance values in the 37 course categories as a vector for each student, then applied a clustering algorithm on the vectors. Since the dimensionality of 37 is too high for most clustering algorithms, we applied PCA to reduce the dimensionality. The clustering algorithm we applied was the DBSCAN clustering algorithm [8] using the Euclidean distance. DBSCAN performs density-based clustering and does not require the input of the cluster number. The parameters we tuned in this experiment are specified in Table 2. We applied silhouette score [21] to select the best set of parameters with the highest silhouette score.

**Table 2. Parameters to tune for the clustering.**

| Parameters | Range |
| --- | --- |
| Number of PCA components | [5, 6, 7, 8, 9, 10, 11, 12, 13, 14, 15] |
| Eps of DBSCAN | [0.1, 0.2, 0.3, 0.4, 0.5, 0.6, 0.7, 0.8, 0.9, 1.0] |
| MinPoints of DBSCAN | [5, 6, 7, 8, 9, 10, 11, 12, 13, 14, 15, 16, 17, 18, 19, 20] |

We applied the same clustering procedure to AR and RAI attendance values respectively. For AR, the procedure failed to generate meaningful clustering (the result contained one big cluster only). For RAI the procedure produced 8 clusters with 616, 472, 665, 520, 130, 1799, 61, and 549 students respectively. 26 student samples were labeled as noise by DBSCAN and excluded in the follow-up study. For each cluster, we identified the top five most popular majors among the samples in the cluster to analyze students' academic interest and performance.

Figure 2 shows the profiles of the 8 clusters from the RAI clustering, labeled as cluster 1 to 8. We combined the dimensions of student numbers, distribution of majors, RAI attendance rates, top 10% academic performance ratio, and last 10% academic performance ratio to show how RAI attendance is related with academic performance and how the analysis can provide guidance on major selection. While some of the findings are interesting, we admit that not all phenomena can be fully explained due to the complexity behind attendance and attainment [15]. For all subfigures in Figure 2, the X is the major of the students. Figure 2(a) shows the number of students in each major for the 8 clusters in a row. Some of the clusters are very specific. Cluster 5 contains two majors only, TRAN (Translation) and PSY (Psychology); Cluster 7 contains the major of FE (Financial Engineering) only. Figure 2(b) shows the distribution of majors among the clusters (whether a cluster accounts for a significant portion of the students in a major), with each bar representing a fraction of the corresponding major in the university. For example, as shown in Figure 2(b), close to 70% of the students majoring in PSY are in cluster 5; close to 50% of the students majoring in CSE (Computer Science and Engineering) are in Cluster 6, with other large portions of CSE students in Cluster 2, 3 and 4. Figure 2(c) shows the RAI attendance of the clusters. Students in Cluster 6 have significantly lower RAI values than the other clusters. Figure 2(d) shows the ratio of students with a GPA in the top 10% of the major. If a bar of major *m* is higher than the 0.1 line, it means that the students from the cluster in major *m* outperform the average level of students in major *m*. Similarly, Figure 2(e) shows the portion of students with a GPA in the last 10% of the major. The higher the value, the worse the performance of the students, which is the opposite of Figure 2(d).

Figure 2 illustrates how RAI correlates with academic performance. Figure 2(c) shows that Cluster 6 has the overall lowest RAI values, with all the five majors having negative RAI values. Cluster 6 also has the worst top 10% ratio in Figure 2(d) (only one major is barely over the average cutline), and the worst last 10% ratio in Figure 2(e) (all five majors worse than the average). The TRAN major has about the same number of students in Cluster 5 and Cluster 8. The TRAN in Cluster 8 has a higher RAI value as well as a higher top 10% ratio and a much lower last 10% ratio than TRAN in Cluster 5. There are exceptions though. CSE in Cluster 3 has a negative RAI, but its top 10% ratio is the highest in Cluster 3. However, this is consistent with our result in Table 1, which shows that CSE courses have a relatively low RAI correlation with academic performance (CSE major students usually take many CSE courses).

Another interesting finding is that in all the 7 clusters with more than one major, the major that has the highest RAI value also has the lowest last 10% ratio except for Cluster 2. In Cluster 2 it is the second highest RAI major EIE that has the lowest last 10% ratio. The highest RAI major in Cluster 2 is BIFC (Bioinformatics), a new major with a relatively small enrollment. Students facing the risk of poor academic performance may consider selecting or switching to the major with the highest RAI in the same cluster. While we admit that this is by-no-mean a correlation between RAI and students' academic interest, we remark that the interest in a subject is generally believed to be a weapon to fight against poor performance. Together with the fact that DBSCAN worked better on RAI than AR, we believe this phenomenon may suggest that RAI has a better potential than AR for exploring students' academic interest.

## 6. LIMITATIONS AND FUTURE WORK

In this study, we defined the Relative Attendance Index to adjust the attendance measurement, with the objective of better reflecting students' attainment and interest. While attendance is affected by many factors [15], the new information we introduced is only the attendance of the peer. Further improvements should address more factors of attendance.

When clustering on student data, the clustering algorithm DBSCAN worked better on RAI than AR data, and the clustering analysis confirmed the correlation between RAI and academic performance. We admit that we have not been able to confirm the correlation between RAI and students' academic interest, which is an interesting topic to be further explored.

The raw attendance data was collected using a WiFi based method [19, 28]. It is possible that some students closed the WiFi connection on their cell phones or even closed their cell phones all together before class, leading to a false label of absence. If a student had less than 50 WiFi connection records in a week, their data in that week were excluded from the statistics. While we admit that this could generate some noise in the attendance, we observed that this situation only occurred rarely. For example, in Fall term 2019 we found that only 2 out of 4838 students turned off WiFi completely, and only 3.34% of the students had some weeks of data filtered out. We conjecture two reasons for this phenomenon. First, usage of laptops and tablets is popular among the students in this university. Many students carry them to the classroom to view course materials. Laptops and tablets usually can connect WiFi only. Secondly, students use time confetti before and after class to chat with friends or read the news. It will be interesting to see how RAI performs on other attendance data such as QR code scanning and online attendance data.

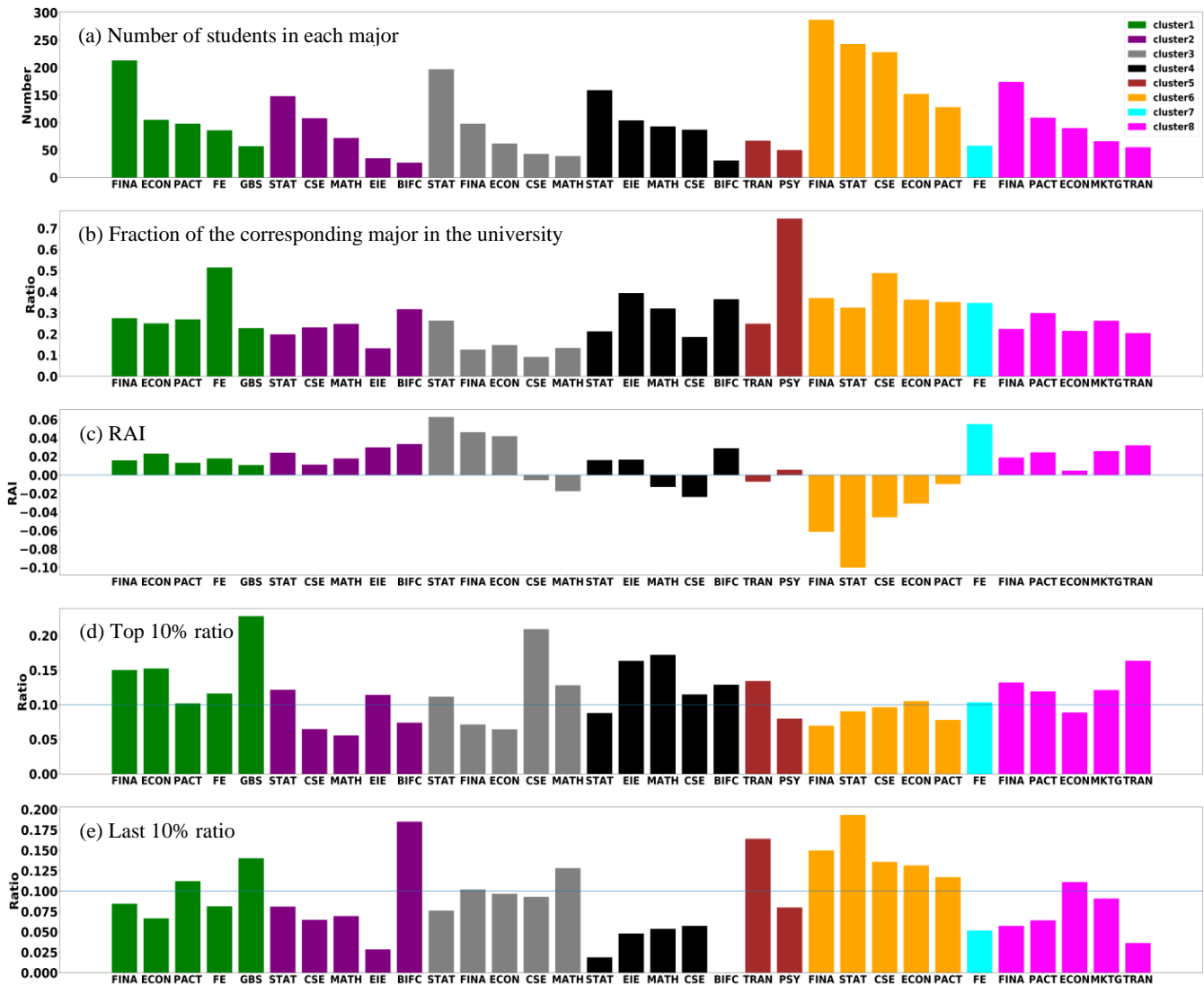

**Figure 2. Student Clustering.**

## 7. ACKNOWLEDGMENTS

This work was supported by Shenzhen Research Institute of Big Data. We also want to thank anonymous reviewers for helpful suggestions.